\documentclass[%
reprint,
amsmath,amssymb,
aps,
]{revtex4-2}
\usepackage{float}
\usepackage{graphicx}
\usepackage{dcolumn}
\usepackage{bm}

\begin{document}
	\preprint{APS/123-QED}
	\title{Quantum metrology of hopping strength in a one-dimensional electronic chain }
	\author{Yuetao Chen$^{1}$}
	\author{Gaiqing Chen $^{1}$}
	\author{jin Wang$^{1}$}
	\author{qiang Ma$^{1}$}
	\author{Shoukang Chang$^{1}$}
	\author{Shaoyan Gao$^{1}$}
	\thanks{Corresponding author. gaosy@xjtu.edu.cn}
	\affiliation{$^{{\small 1}}$\textit{MOE Key Laboratory for Nonequilibrium Synthesis and
			Modulation of Condensed Matter, Shaanxi Province Key Laboratory of Quantum
			Information and Quantum Optoelectronic Devices, School of Physics, Xi'an
			Jiaotong University, 710049, China}}
	
	\begin{abstract}
		The electron hopping between the two sites in a lattice is of fundamental importance in condensed matter physics. Precise control of the hopping strength allows for the prospect of manipulating the properties of electronic materials, such as topological properties, superconductivity, etc. In this framework, measuring the hopping strength of an electronic lattice with high precision is perhaps the most relevant step in controlling the properties of electronic materials. Here, we design a critical quantum metrological protocol to measure the hopping strength in a cavity electronic chain coupling system featuring a pseudo-superradiant phase transition. We show that the cavity ground state, which is initially a squeezed vacuum state, can be utilized as a quantum probe to achieve a high quantum precision of the hopping strength, which can be optimally saturated in either the loss or lossless case. Remarkably, in the presence of chain loss, we find that increasing the electron current in the chain is beneficial for enhancing precision, and the arbitrarily large precision could be obtained by increasing the chain size, in principle. Our results provide an effective method to measure the hopping strength in the electronic chain with high precision, so it has potential applications in critical quantum metrology, condensed matter physics, etc.

	\end{abstract}
	
	\maketitle
	
	\section{Introduction}
	 Hopping strength, as an important physical parameter describing the electron hopping behavior within a crystal, plays a crucial role in the formation of the electronic band structure, thereby affecting the properties of electronic materials. Some phase transitions, such as the topological phase transition in a Su-Schrieffer-Heeger chain (also known as a prototypical model), parity-time ($\mathcal{PT}$)-symmetry phase transition in a Hatano-Nelson chain, etc., need highly precise control of hopping strength near the phase transition point. Thus, it is crucial to realize the measurement of hopping strength with high precision in a solid electronic material.
	
	Critical quantum metrology has received unprecedented attention recently following the realization of high-precision measurement by exploiting quantum fluctuations in the proximity of the criticality. This long-standing
	idea has been recently considered in the quantum regime by
	exploiting quantum phase transitions in the ground state or
	dissipative steady states \cite{dissipativesteadystates1,dissipativesteadystates2,dissipativesteadystates3}, of many-body \cite{manybody1,manybody2,manybody3,manybody4} or
	light-matter interacting systems \cite{lightmatterinteractingsystems1,lightmatterinteractingsystems2,lightmatterinteractingsystems3,lightmatterinteractingsystems4}. In a typical critical quantum metrology protocol \cite{groundstate1,PRL7,manybody2,manybody1}, the parameter to be estimated is encoded in the system ground state of the lossless Hamiltonian, or the system steady state in the presence of a dissipative process. Close to the phase transition point, the susceptibility of the equilibrium state becomes infinite, indicating that it is possible to obtain extremely precise estimations. 
	
	Such phase transitions required for critical quantum metrology can occur in condensed matter cavity QED systems (also known as quantum Floquet engineering), in which a cavity is coupled to an electronic material to inhibit the heating effect in conventional Floquet engineering \cite{crossover}. For instance, a superradiant phase transition induced by electron-electron interactions \cite{superradiantphasetransition}, a $\mathcal{PT}$-symmetry phase transition in a cavity non-Hermitian electronic lattice interaction system \cite{PTtransition}, a photon condensation phase transition in magnetic cavities \cite{photoncondense}, etc. In these systems, the ground state of the cavity will experience a phase transition which depends on the hopping strength in the electronic material. This will result in divergent susceptibilities, which could be exploited for highly precise parameter estimation.
	
\begin{figure}[h]
		\label{Fig.1} \centering \includegraphics[width=0.95\columnwidth]{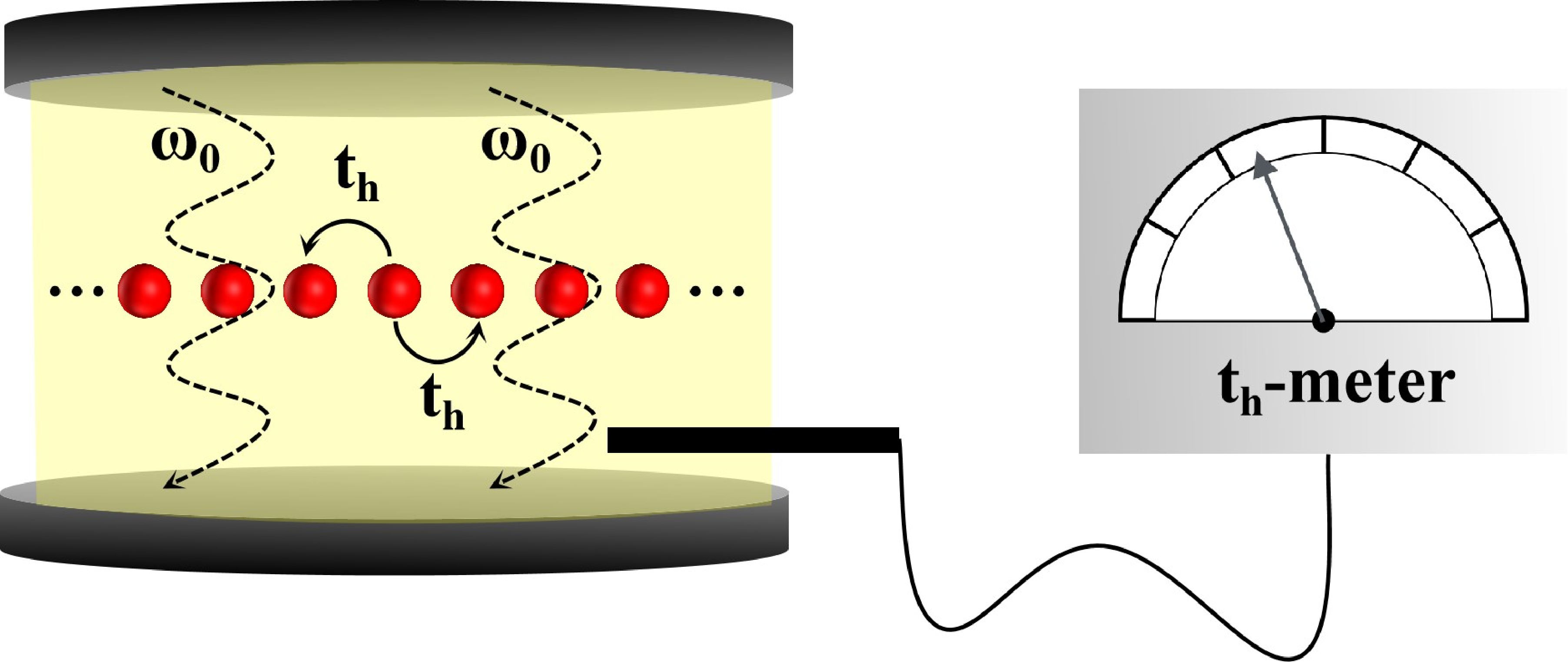}%
		\caption{Sketch of the investigated setup: A tight-binding ODC with nearest neighbor hopping strength $t_{h}$ is coupled to a single mode cavity with frequency $\omega_{0}$. The parameter $t_{h}$ can be estimated experimentally by performing an appropriate measurement on the coupled light-matter system.}
	\end{figure}

	Motivated by the potential use of critical quantum metrology in estimating  parameters in electronic materials, in this paper, we design parameter-estimation protocols based on equilibrium properties. In particular, we investigate a tight-binding one-dimensional chain (ODC) for a solid electronic material coupled to the quantized light field of a cavity, in which a pseudo superradiant phase transition can occur. The ODC is coupled to the cavity through a quantized version of the Peierls substitution. By switching on the interaction between the ODC and the cavity, hopping strength information may be mapped to the ground state of the cavity and extracted through optical detection (see Fig. \ref{Fig.1}). We find analytical expressions for the quantum Fisher information(QFI) of the hopping strength, and the classical Fisher information(CFI) is also calculated analytically by using homodyne detection. We have three main results: (i) the quantum precision of the hopping strength, limited by the QFI, is extremely large near the critical point and can be completely saturated by using homodyne detection in the ideal case; (ii) in the presence of cavity dissipation, the quantum precision limit can be nearly saturated given that the photon decay $\kappa_{ph}<0.3$. This can be achieved in the recent condensed matter cavity QED experiments; and (iii) by controlling the two dissipative processes in the ODC and increasing the ODC size, one can obtain the arbitrarily large quantum precision of the hopping strength in principle. It can still be nearly saturated in the case of the small photon decay rate and large ODC size. Our results are helpful for fields like critical quantum metrology, condensed matter physics, and condensed matter cavity QED. 
	
	This paper is organized as follows: In Sec. \ref{ODC-cavity interaction model}, we introduce the ODC-cavity interaction model and the pseudo superradiant phase transition in this model. In Sec. \ref{Quantum parameter estimation}, we briefly introduce the theory of quantum parameter estimation and homodyne detection. Meanwhile, the analytical expressions of the QFI and CFI are derived, and we study them. Sec. \ref{Cavity Dissipation} accounts for the study of the QFI and CFI in the case of cavity dissipation. In Sec. \ref{ODC Dissipation}, we study the dynamics of the cavity field in the presence of ODC dissipation. At the same time, the analytical expressions of the QFI and CFI are derived and investigated.  Finally, we present the conclusions of our results in Sec. \ref{CONCLUSION}.
	
	\section{RESULTS AND DISCUSSION}
	\subsection{ODC-cavity interaction model}
	\label{ODC-cavity interaction model}
	We consider a one-dimensional chain coupled to a single mode cavity with frequency $\omega_{0}$ as depicted in Fig. 1. The corresponding Hamiltonian reads
	\begin{eqnarray}
		H&=&-\sum_{j=1}^{L}\left[t_{h}e^{-i\frac{g}{\sqrt{L}}(\hat{a}^{\dagger}+\hat{a})}\hat{c}_{j+1}^{\dagger}\hat{c}_{j}+\mathrm{h.c.}\right]\notag\\
		&&+\omega_{0}\left(\hat{a}^{\dagger}\hat{a}+\frac{1}{2}\right),\label{1}
	\end{eqnarray}
	where $\omega_{0}$ is the frequency of the cavity mode, $\hat{a}$ and $\hat{a}^{\dagger }$ are bosonic creation and annihilation operators of the cavity mode, $\hat{c_{j}}$ and $\hat{c}^{\dagger}_{j}$ are the fermionic annihilation and creation operators at lattice cite $j$, and $L$ is the number of lattice cites. $\hat{a}$ and $\hat{a}^{\dagger}$ are related to the quantized electromagnetic vector potential through $A=\frac{g}{\sqrt L}(\hat{a}+\hat{a}^{\dagger})$, with $e=\hbar=c=1$. We consider periodic boundary conditions and set the lattice constant to 1. The coupling strength $g$ depends on the geometry and material composition of the cavity. In momentum space, the Hamiltonian can be rewriten as 
	\begin{eqnarray}
		H&=&\cos\left(\frac{g}{\sqrt{L}}(\hat{a}+\hat{a}^\dagger)\right)\hat{T}+\sin\left(\frac{g}{\sqrt{L}}(\hat{a}+\hat{a}^\dagger)\right)\hat{J}\notag\\
		&&+\omega_0(\hat{a}^\dagger \hat{a}+\frac{1}{2}),\label{2}
	\end{eqnarray}
	where
	\begin{eqnarray}
		\hat{T}&:=&\sum_{k}-2t_{h}\cos(k)\hat{c}_{k}^{\dagger}\hat{c}_{k}\notag\\
		\hat{J}&:=&\sum_{k}2t_{h}\sin(k)\hat{c}_{k}^{\dagger}\hat{c}_{k}\label{3}
	\end{eqnarray}
	\begin{figure*}[ht]
		\label{Fig.2} \centering \includegraphics[width=1.5\columnwidth]{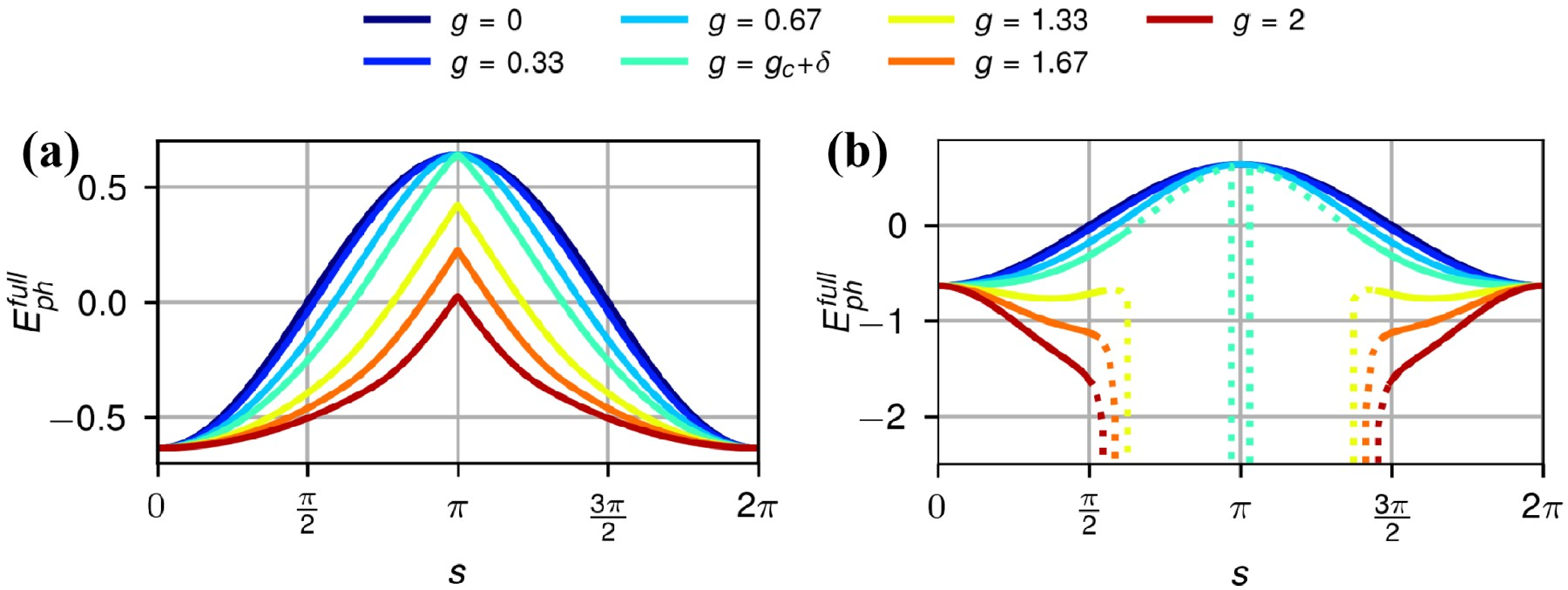}%
		\caption{(a)Spectrum of the full cavity Hamiltonian as function of the center of the Fermi sea (FS), with the electronic state chosen as a single connected momentum region being occupied (Fermi sea, FS). (b)Spectrum of the the second-order truncated cavity Hamiltonian as function of the center of the Fermi sea (FS). The parameters are chosen as $\omega_{0}=t_h=1$ and $L=400$.}
	\end{figure*}
	are the kinetic energy and current operators, respectively. $\hat{c_{k}}$ and $\hat{c}^{\dagger}_{k}$ are the fermionic annihilation and creation operators at momentum $k$. We will analyze quantum-metrology protocols to estimate hopping strength $t_{h}$, assuming in each case that all other parameters are known and there is no entanglement between photons and electrons.
	
	In order to evaluate the performance of these protocols, we first need to characterize the system ground state as a function of the system parameters.  
	The electron ground state is the Fermi sea (FS) centered at $s$ (see Appendix A).  After considering the half-filled condition,  the center is located at $s = 0$.  In Fig. 2a, we illustrate this by showing the spectrum of the full cavity Hamiltonian (Eq. (\ref{A3}) in Appendix A) as a function of the FS center for fixed parameters with $\omega_{0}=t_h=1$ and $L=400$. The minimum cavity energy is bounded and always located at the FS centered around $s = 0$ for all considered coupling strengths. This reveals that the electron ground state remains unchanged when the ODC-cavity interaction occurs. In this electron ground state, the expectation value of the current density operator $\bar{\hat{J}}=0$ (see Appendix A).
	
	At finite $L$, the Hamiltonian can be truncated at second order in $A=\frac{g}{\sqrt L}(\hat{a}+\hat{a}^{\dagger})$
	\begin{eqnarray}
		H^{2^{\mathrm{nd}}}&=&T+\frac{g}{\sqrt{L}}\left(\hat{a}^{\dagger}+\hat{a}\right)\hat{J}-\frac{1}{2}\frac{g^{2}}{L}\left(\hat{a}^{\dagger}+\hat{a}\right)^{2}\hat{T}\notag\\
		&&+\omega_{0}\left(\hat{a}^{\dagger}\hat{a}+\frac{1}{2}\right) \label{4}
	\end{eqnarray}
	The diamagnetic term ${\frac{1}{2}}{\frac{g^{2}}{L}}\left(\hat{a}^{\dagger}+\hat{a}\right)^{2}\hat{T}$ yields a phase transition towards the pseudo-superradiant phase at $g_{c}=\sqrt{\frac{\pi\omega_{0}}{4t_{h}}}$. However, as $g>g_{c}$, this truncation breakdown is due to the absence of a photon ground state. We illustrate this by showing the spectrum of the second-order truncated cavity Hamiltonian as a function of the FS center in Fig. 2(b). The minimum cavity energy is not located at the FS centered around $s = 0$ anymore as $g>g_{c}$. At some point (dotted line), the effective cavity frequency vanishes, leading to the absence of a photon ground state. Thus, the second-order truncated cavity Hamiltonian is valid as $g<g_{c}$, and it can be diagonalized using a squeezing transformation. The photon ground state is given by $|\psi_{ph}\rangle=S^\mathrm{sq}[\bar{\hat{T}}]|0\rangle$, while $S^\mathrm{sq}[\hat{T}]=\frac14\ln\left(\frac{\mathcal{V}[\hat{T}]}{\omega_0}\right)\left(\hat{a}^2-(\hat{a}^\dagger)^2\right)$ is a squeezing operator with $\mathcal{V}[\hat{T}]=\omega_0\sqrt{1-2\frac{g^2}{L\omega_0}\hat{T}}$.  
	\subsection{Quantum parameter estimation}
	\label{Quantum parameter estimation}
	We are interested in the precise estimation of $t_{h}$ which is bounded
	by the quantum Cramer-Rao (CR) bound: $\delta^{2}t_{h}\geq(\nu\mathcal{I}_{t_{h}})^{-1}$, where $\nu$ is the number of system copies and $\mathcal{I}_{t_{h}}$ is the quantum Fisher information of the parameter $t_{h}$. We can use the photon ground state to calculate the quantum Fisher information (QFI) $\mathcal{I}_{t_{h}}=4[\langle\partial_{t_{h}}\psi_{ph}|\partial_{t_{h}}\psi_{ph}\rangle+(\langle\partial_{t_{h}}\psi_{ph}|\psi_{ph}\rangle)^{2}]$. After straightforward calculation, the analytical expression of $\mathcal{I}_{t_{h}}$ reads
	\begin{equation}
		\mathcal{I}_{t_{h}}=2\left(\frac{g^2}{2\pi\omega_{0}\left(1-\frac{4g^2t_{h}}{\pi \omega_{0}}\right)}\right)^2, \label{5}
	\end{equation}
	which shows the divergence of $\mathcal{I}_{t_{h}}$ at the critical point $g=g_{c}$. It means that an extreme large estimation precision of $t_{h}$ could be obtained near the critical point of coupling strength.This is consistent with previous investigations on metrology based on phase transitions in light-matter systems \cite{PRL7,PRL27,Criticalparametricquantumsensing,CriticalQuantumMetrology}. The bound given by ${I}_{t_{h}}$ sets the ultimate precision limit allowed by quantum mechanics. However, the best measurement is not clearly provided by the quantum Cram\'{e}r-Rao theorem. To saturate the bound, one needs to implement the optimal positive-operator-valued measure (POVM). In fact, there are many distinct types of detection methods for parameter measurement, such as homodyne detection \cite{homodynedetection1,homodynedetection2}, intensity detection \cite{intensitydetection1,intensitydetection2}, and parity detection \cite%
	{paritydetection1}. Moreover, compared with intensity detection and parity
	detection, homodyne detection is easily achieved using existing experimental
	technology. Thus, we evaluate the classical Fisher information (CFI) using homodyne detection
	on the cavity field. This detection can be performed along any quadrature $\hat{x}_{\phi} = \operatorname{cos}\left(\phi\right)\hat{x} + \operatorname{sin}\left(\phi\right)\hat{p}$ with local oscillator phase $\phi$. The classical Fisher information $F_{{t_{h}}}(\phi)$ associated with several choices of quadratures can be calculated by 
	\begin{align}
		F_{{t_{h}}}(\phi)&=\int dx\frac{\left[{\partial p_{\phi}(x)}/{\partial t_{h}}\right]^2}{p_{\phi}(x)},\notag\\
		p_\phi(x)&=\int dy W[\rho_{ph}](x\cos\phi-y\sin\phi,x\sin\phi+y\cos\phi), \label{6}
	\end{align}
	where $W[\rho_{ph}]$ and $\rho_{ph}$ are the Wigner function and the density matrix of the photon state in the cavity, respectively. After straightforward calculation, the analytical expression of $F_{{t_{h}}}(\phi)$ read
	\begin{align}
		F_{{t_{h}}}(\phi)=\frac{\pi^2\omega_0^2\left(\left(1-\frac{4g^{2}t_h}{\pi\omega_0}\right)\frac{2g^2\cos[\phi]^2}{\pi\omega_0}-\frac{g^2\sin[\phi]^2}{\pi\omega_0}\right)^2}{4g^4\left(\left(1-\frac{4g^{2}t_h}{\pi\omega_0}\right)\cos[\phi]^2+\sin[\phi]^2\right)^2}. \label{7} 
	\end{align}
	\begin{figure*}[ht]
		\label{Fig.3} \centering \includegraphics[width=2\columnwidth]{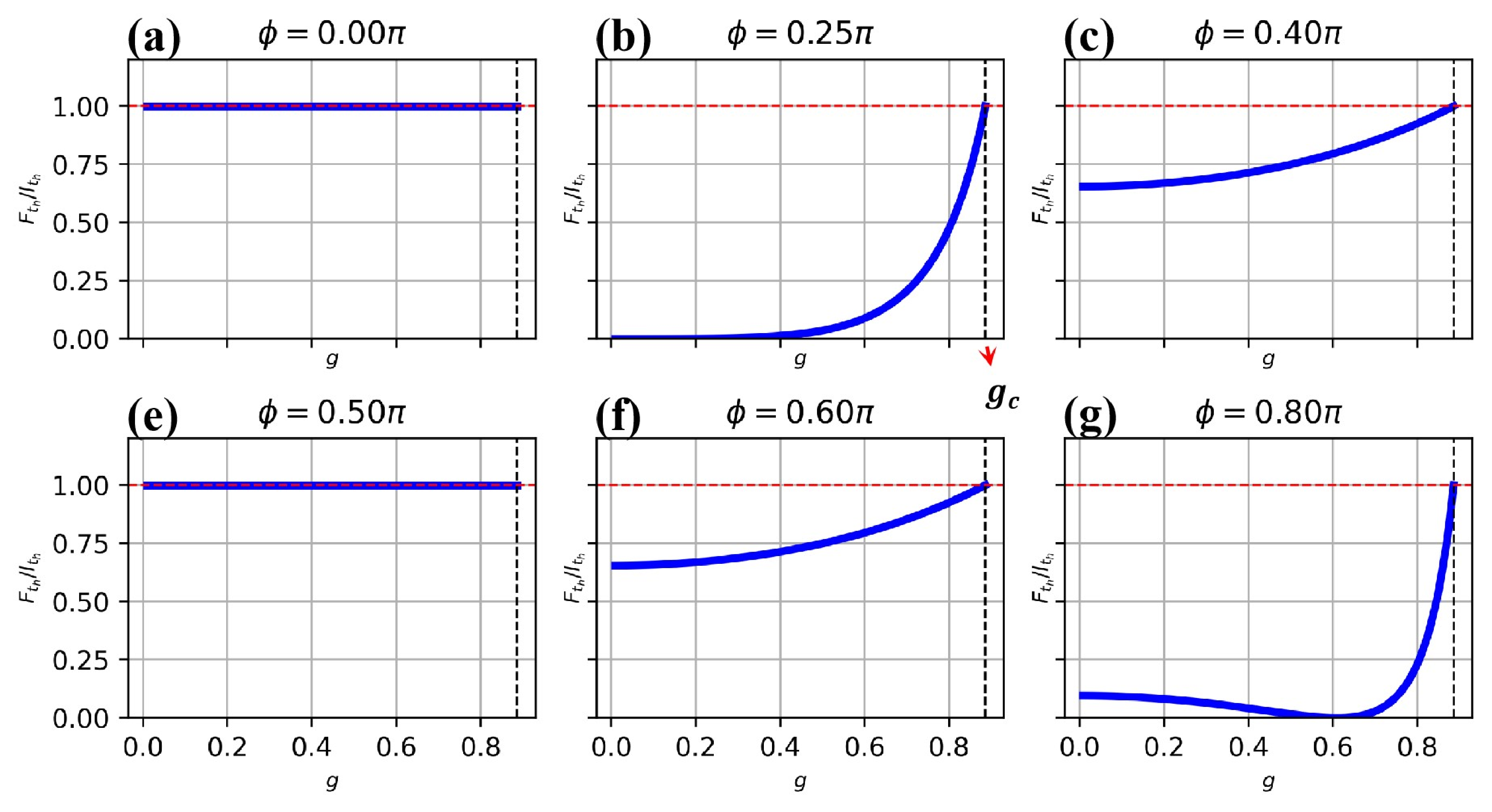}%
		\caption{$F_{{t_{h}}}(\phi)$-to-$\mathcal{I}_{t_{h}}$ ratio versus coupling strength $g$, for a homodyne measurement performed along various quadratures $\hat{x}_{\phi}$. (a)$\phi=0$, (b)$\phi=0.25\pi$, (c)$\phi=0.4\pi$, (d)$\phi=0.5\pi$, (e)$\phi=0.6\pi$, (f)$\phi=0.8\pi$. Black dashed line represents the critical point $g_{c}=\sqrt{\frac{\pi\omega_{0}}{4t_{h}}}$ and red dashed line represents $F_{{t_{h}}}(\phi)/\mathcal{I}_{t_{h}}=1$. The parameters are the same as those in Fig. 2.}
	\end{figure*}
	It is worth emphasizing that both QFI and CFI are independent of the number of lattice cites $L$, according to Eqs. (\ref{6}) and (\ref{8}). This reveals that quantum parameter estimation in an electronic chain can be performed without the extremely large chain size. To quantify the performance of the quantum-metrology protocol, in Fig. 3, we show the ratio $F_{{t_{h}}}(\phi)/\mathcal{I}_{t_{h}}$ for homodyne detection performed on the chosen quadrature. In general, the ratio depends on $g$, and it becomes optimal near the critical point. For $\phi=0$ and $\phi=0.5\pi$, the optimal precision is reached for all values of $g<g_c$, hence the quantum Fisher information is saturated by homodyne measurement along the quadrature $\hat{x}$ or $\hat{p}$.
	\subsection{Cavity Dissipation}
	\label{Cavity Dissipation}
	The above results are valid for isolated quantum systems. However, quantum systems are inevitably prone to decoherence, which makes them especially fragile and generally reduces the performance of quantum-metrology protocols when they interact with an environment. Therefore, an open quantum system dynamic are needed to explain the evolution of the system. Firstly, we only consider cavity dissipation, assuming that the ODC will not experience any dissipative processes. To this extent, we investigate the ODC-cavity master equation, given in the Lindbladian form of 
	\begin{align}
		\partial_t\hat{\rho}=-i{\left[H^{2^{\mathrm{nd}}},\hat{\rho}\right]}+\kappa_{ph}\mathcal{L}{\left[\hat{a}\right]}\hat{\rho}, \label{8} 
	\end{align}
	where $\hat{\rho}=\hat\rho_{ph}\otimes\hat\rho_{el}$ is the density matrix of ODC-cavity system, $\kappa_{ph}$ is the photon decay rate, and the Lindbladian superoperator term is denoted by $\mathcal{L}[\hat{O}]\hat{X}=2\hat{O}\hat{X}\hat{O}^{\dagger}-\hat{X}\hat{O}^{\dagger}\hat{O}-\hat{O}^{\dagger}\hat{O}\hat{X}$.
	Notice that we are considering a reduced master equation as we mainly focus on the fate of the photonic field. In order to characterize the dynamics of the photonic field, we will reduce the electronic degrees of freedom with the assumption that the ODC and cavity are in their ground states at the beginning. Details can be found in Appendix B. This yields an effective master equation for the state of the photonic field:
	\begin{equation}
		\partial_{t}\hat{\rho}_{ph}=-i\Bigg[ \omega_{0}\hat{a}^{\dagger}\hat{a}-\omega_{0}X \frac{g^{2}}{4}(\hat{a}+\hat{a}^{\dagger})^{2},\hat{\rho}_{ph}\Bigg]+\kappa_{ph}\mathcal{L}[\hat{a}]\hat{\rho}_{ph}, \label{9}
	\end{equation}
	where we defined $X=\frac{4t_h}{\omega_0\pi}$. Notice that we calculate the  expectation values of the kinetic energy $\hat{T}$ and current operators $\hat{J}$ in Eq. (\ref{10}), and they remain unchanged in the ground state of the electron due to the fact that $[\hat{c}_k^\dagger \hat{c}_k,H^{2^{\mathrm{nd}}}]=0$. Since this equation is quadratic in $\hat{a}$, it can be solved by a Gaussian ansatz. The dynamics of the photonic field are then fully featured by the evolution equation for the covariance matrix $\sigma$ of Gaussian states, while the displacement vector is damped to zero quickly and can be quit safely. Therefore, we obtain $\partial_{t}\sigma=B\sigma+\sigma B^{T}-2\kappa_{ph}(\sigma-\sigma^{Lin})$, where 
	\begin{equation}
		B=\begin{bmatrix}0&\omega_0\\\omega_0(Xg^2-1)&0\end{bmatrix},
		\sigma^{ Lin}=\begin{bmatrix}1/2&0\\0&1/2\end{bmatrix} \label{10}
	\end{equation}
	We find the steady state by solving $\partial_{t}\sigma=0$. 
	\begin{figure*}[ht]
		\label{Fig.4} \centering \includegraphics[width=1.6\columnwidth]{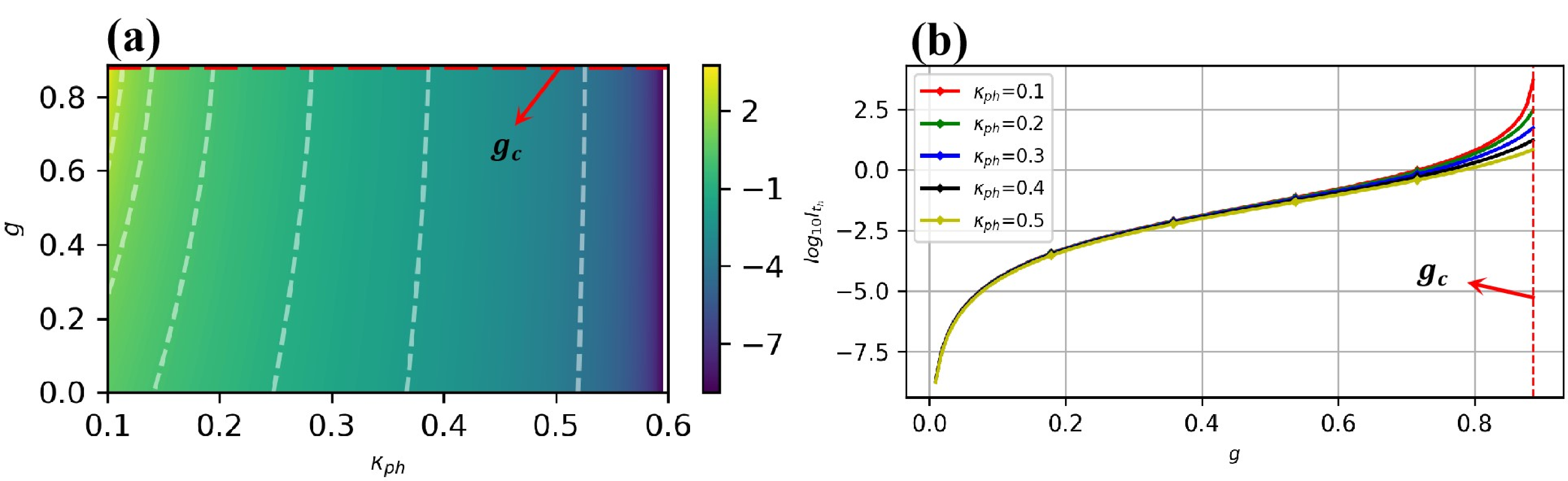}%
		\caption{(a) The QFI $log_{10}\mathcal{I}_{t_h}$ as functions
			of $\kappa_{ph}$ and $g$. In (b), we show the QFI $log_{10}\mathcal{I}_{t_h}$ as functions of $g$ for some considered $\kappa_{ph}$ values. The red dashed line denotes the critical point $g=g_c$. The other parameters are chosen as $\omega_{0}=t_h=1$ and $L=400$.}
	\end{figure*}
	The steady state is a squeezed thermal state, with $\sigma$ is given by
		\begin{equation}
			\sigma=\begin{bmatrix}\frac{2\kappa_{ph}^2+\omega_0^2-\omega_0^2(Xg^2-1)}{4(\kappa_{ph}^2-\omega_0^2(Xg^2-1))}&\frac{Xg^2\kappa_{ph}\omega_0}{4(\kappa_{ph}^2-\omega_0^2(Xg^2-1))}\\\frac{Xg^2\kappa_{ph}\omega_0}{4(\kappa_{ph}^2-\omega_0^2(Xg^2-1))}&\frac{2\kappa_{ph}^2+\omega_0^2(Xg^2-1)^2-\omega_0^2(Xg^2-1)}{4(\kappa_{ph}^2-\omega_0^2(Xg^2-1))}\end{bmatrix}, \label{11}
		\end{equation}
	where the phase transition point is unshifted $g=g_c$ compared to \cite{Criticalparametricquantumsensing}. Since this state is Gaussian and its first-moment vector is zero, the QFI can be calculated as \cite{arxiv1} (dots denote derivative with respect to the parameter $t_h$)
	\begin{eqnarray}
	\mathcal{I}_{t_h}&=&\frac{128}{256\mathrm{Det}^{2}\sigma-1}\bigg\{\mathrm{Det}^{2}\sigma\mathrm{Tr}[(\sigma^{-1}\dot{\sigma})^2]-\mathrm{Tr}[(\dot{\sigma}M)^2]\bigg\},\notag\\
	M&=&\begin{bmatrix}0&1/2\\[0.3em]-1/2&0\end{bmatrix} , \label{12}
	\end{eqnarray}
	
After straightforward calculation, the analytical expression of the QFI for the estimation of $t_h$ are given by
	\begin{widetext}
		\begin{equation}
			\begin{split}
			\mathcal{I}_{t_h}=\frac{{g}^{4} \omega_{0}^{2} \left(6 \mathrm{A}^{3}+{g}^{2} \omega_{0}^{2} \mathrm{X} \left(-12 \mathrm{A}^{2}+6 {g}^{2} \mathrm{A} \left(\kappa_{{ph}}^{2}+2 \omega_{0}^{2}\right) \mathrm{X}-6 {g}^{4} \omega_{0}^{2} \mathrm{A} \mathrm{X}^{2}+{g}^{6} \omega_{0}^{4} \mathrm{X}^{3}\right) \right) \mathrm{X}^{2}}{2 \left(\mathrm{A}-{g}^{2} \omega_{0}^{2} \mathrm{X}\right)^{2} \left(5 \mathrm{A}+{g}^{2} \omega_{0}^{2} \mathrm{X} \left({g}^{2} \mathrm{X}-5\right)\right) \left(3 \mathrm{A}+{g}^{2} \omega_{0}^{2} \mathrm{X} \left({g}^{2} \mathrm{X}-3\right)\right) {t}_{{h}}^{2}}, \label{13}
			\end{split}
		\end{equation}
	\end{widetext}
	where $\mathrm{A}=\kappa_{\mathrm{ph}}^{2}+\omega_{0}^{2}$. Generally speaking, the cavity loss is not beneficial to the performance of quantum-metrology protocols. We illustrate this by showing the QFI $log_{10}\mathcal{I}_{t_h}$ as a function of the photon decay rate $\kappa_{ph}$ and the coupling strength $g$ in Fig. 4(a). As the figure shows, the QFI decreases as the photon decay rate increases. Furthermore, with the fixed photon decay rate, large QFI can be obtained near the critical point $g=g_c$, as shown in Fig. 4(b). This indicates that one can always adjust the set of coupling strengths $g$ close to the critical point to deliver maximal QFI of $t_h$ in the dissipative cavity. 

	To quantify the performance of quantum-metrology protocols in estimating hopping strength $t_h$ in the cavity dissipative process, we evaluate the CFI $F_{{t_{h}}}(\phi)$ using homodyne detection performed on the quadrature. By using Eq. \ref{7} again, one can obtain the CFI. Notice that the Wigner function of the photon state in the cavity $W[\rho_{ph}]$ has been obtained by finding the steady photon state in Appendix B. 
	After some straightforward calculations, the analytical expression of the CFI for the estimation of $t_h$ is given by 
	\begin{widetext}
	\begin{equation}
		F_{{t_{h}}}(\phi)=\frac{{g}^{4}\omega_{0}^{2}\left(\mathrm{A}\left(\omega_{0}\cos\left(2\phi\right)+\kappa_{{ph}}\sin\left(2\phi\right)\right)+{g}^{2}\omega_{0}\sin\left(\phi\right)^{2}\left(2\mathrm{A}-{g}^{2}\omega_{0}^{2}\mathrm{X}\right)\right)^{2}\mathrm{X}^{2}}{\left(\mathrm{A}-{g}^{2}\omega_{0}^{2}\mathrm{X}\right)^{2}\left(2\mathrm{A}+{g}^{2}\omega_{0}\mathrm{X}\left(\omega_{0}\left(\cos\left(2\phi\right)-2\right)+\kappa_{{ph}}\sin\left(2\phi\right)+{g}^{2}\omega_{0}\sin\left(\phi\right)^{2}\mathrm{X}\right)\right)^{2}{t}_{{h}}^{2}}. \label{14}
	\end{equation}
	\end{widetext}
	\begin{figure*}[ht]
		\label{Fig.5} \centering \includegraphics[width=1\columnwidth]{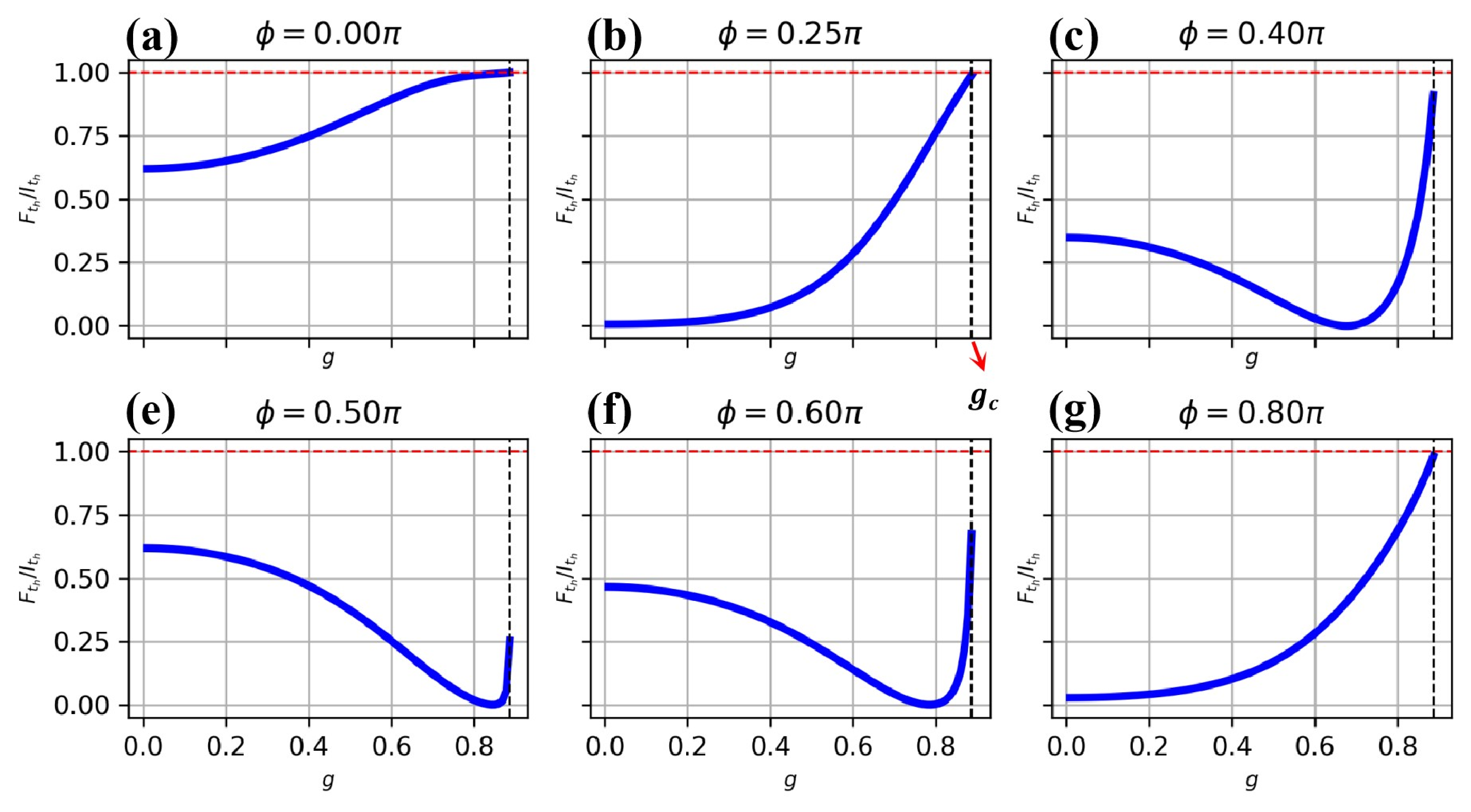}%
		\caption{$F_{{t_{h}}}(\phi)$-to-$\mathcal{I}_{t_{h}}$ ratio versus coupling strength $g$, for a homodyne measurement performed along various quadratures $\hat{x}_{\phi}$. (a)$\phi=0$, (b)$\phi=0.25\pi$, (c)$\phi=0.4\pi$, (d)$\phi=0.5\pi$, (e)$\phi=0.6\pi$, (f)$\phi=0.8\pi$. Black dashed line represents the critical point $g_{c}=\sqrt{\frac{\pi\omega_{0}}{4t_{h}}}$ and red dashed line represents $F_{{t_{h}}}(\phi)/\mathcal{I}_{t_{h}}=1$. $\kappa_{ph}$ is 0.1 and other parameters are the same as those in Fig. 2.}
	\end{figure*}
	In Fig. 5, we show the ratio $F_{{t_{h}}}(\phi)/\mathcal{I}_{t_{h}}$ for homodyne detection performed on the chosen quadrature in the cavity dissipative condition. Compared to the ideal cases, optimal precision cannot be reached for all considered $\phi$ when $g$ is far away from the critical point $g_c$. For $\phi=0$, $\phi=0.25\pi$, and $\phi=0.8\pi$, the QFI can be saturated by homodyne measurement at the critical point $g=g_c$. Furthermore, the ratio becomes optimal in proximity of the critical point for $\phi=0$. This reveals that the best choices of quadrature are $\hat{x}_{\phi}=\hat{x}_{0}=\hat{x}$.
	It is also important to see how the ratio $F_{{t_{h}}}(\phi)/\mathcal{I}_{t_{h}}$ is affected by the presence of cavity dissipation. To investigate this, we show the ratio $F_{{t_{h}}}(\phi)/\mathcal{I}_{t_{h}}$ as a function of the photon decay rate $\kappa_{ph}$. As is evident from the figure, the ratio is nearly 1 when the photon decay rate $\kappa_{ph}$ is weak, and it is attenuated as $\kappa_{ph}$ increases. However, for values up to
	$\kappa_{ph}=0.3$, one can still obtain a near-optimal ratio of 0.95. This indicates that homodyne detection remains the optimal POVM, as long as $\kappa_{ph}<0.3$. Furthermore, the ratio shows saturation up to 0.9 even in the presence of a photon decay rate of $\kappa_{ph}=0.4$. For the condition of decay rates $\kappa_{ph}>0.55$, the ratio drops faster, although even for
	the extreme decay rate of $\kappa_{ph}=0.7$, the ratio still reaches values
	up to 0.65. In ODC-cavity systems, it is generally conceded that
	the cavity photon decay rate, quantified by $\kappa_{ph}$ in Eq. (\ref{9}),
	from the cavity is the main source of cavity dissipation. Thanks to recent experimental achievements, a terahertz defect cavity with a quality factor up to $10^3$, i.e., $\kappa_{ph}=0.001$, has been used to investigate condensed matter cavity QED \cite{cavitylossexperiment1}. Thus, homodyne detection may be an optimal POVM to estimate the hopping strength in an ODC cavity system based on a condensed matter cavity QED platform. 
	\begin{figure}[ht]
		\label{Fig.6} \centering \includegraphics[width=0.9\columnwidth]{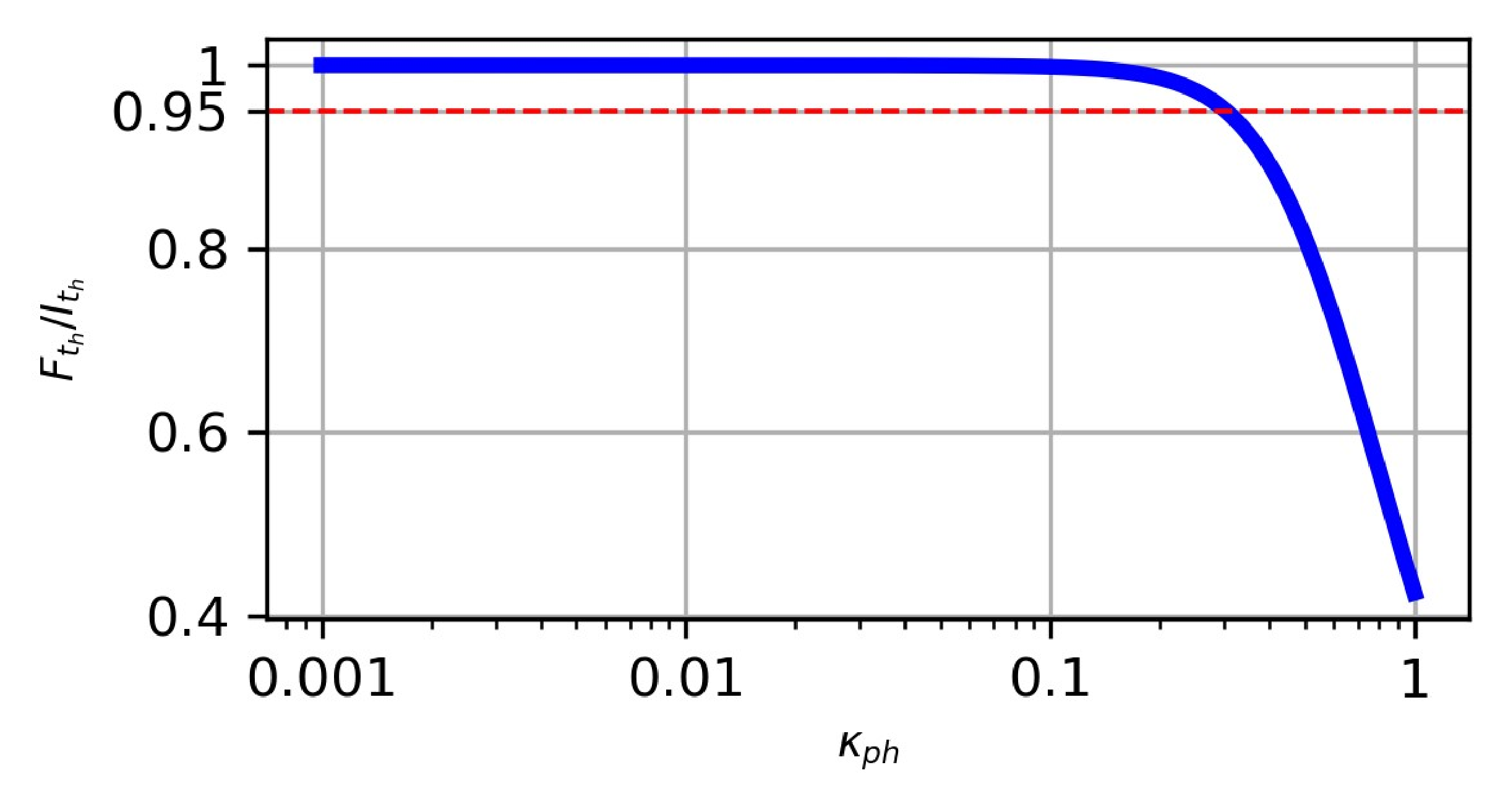}%
		\caption{$F_{{t_{h}}}(\phi)$-to-$\mathcal{I}_{t_{h}}$ ratio versus the photon decay rate $\kappa_{ph}$, for a homodyne measurement performed along the quadratures $\hat{x}$. Red dashed line represents the value of 0.95. The coupling strength $g=0.88$, and other parameters are the same as those in Fig. 2.}
	\end{figure}   
	\subsection{ODC Dissipation}
	\label{ODC Dissipation}
	The above results are valid for quantum systems where the photon in a cavity decays to the outside. However, recent studies \cite{nonhermitian1,nonhermitian2} show that the nonreciprocal hopping that results in non-hermiticity in ODC can be viewed as a dissipative process. Thus, it is also important to study the effect of the ODC dissipative process on the performance of quantum-metrology protocols. In order to characterize
	the ODC dissipative case, we will generalize the results obtained above by considering the ODC dissipative process in the master equation: $\partial_t\hat{\rho}=-i{\left[H^{2^{\mathrm{nd}}},\hat{\rho}\right]}+\kappa_{ph}\mathcal{L}{\left[\hat{a}\right]}\hat{\rho}-2\kappa_{el}\left(1+\sin(k)\right)\sum_k\mathcal{D}{\left[\hat{c}_k\right]}\hat{\rho}-2\Gamma\sum_k\mathcal{D}{\left[\hat{c}_k^{\dagger}\right]}\hat{\rho}$, where $\kappa_{el}$ and $\Gamma$ are the electronic
	decay and pumping rate. Notice that we have rewritten the master equation in the momentum space by considering the periodical condition and assuming that the cavity is not coupled to the ODC dissipative process (see Appendix C). Then, we will reduce the electronic degrees of freedom to study the dynamics of the photonic field (see Appendix D). This yields an effective master equation for the state of the photon field:  
	\begin{equation}
		\partial_t\hat{\rho}_{ph}=-i\biggl[{H}^{2^{\mathrm{nd}}}(\overline{\hat{T}}(t),\overline{\hat{J}}(t)),\hat{\rho}_{ph}\biggr]+\kappa_{ph}L\bigl[\hat{a}\bigr]\hat{\rho}_{ph},\label{15}
	\end{equation}
	  \begin{figure*}[ht]
		\label{Fig.7} \centering \includegraphics[width=1.8\columnwidth]{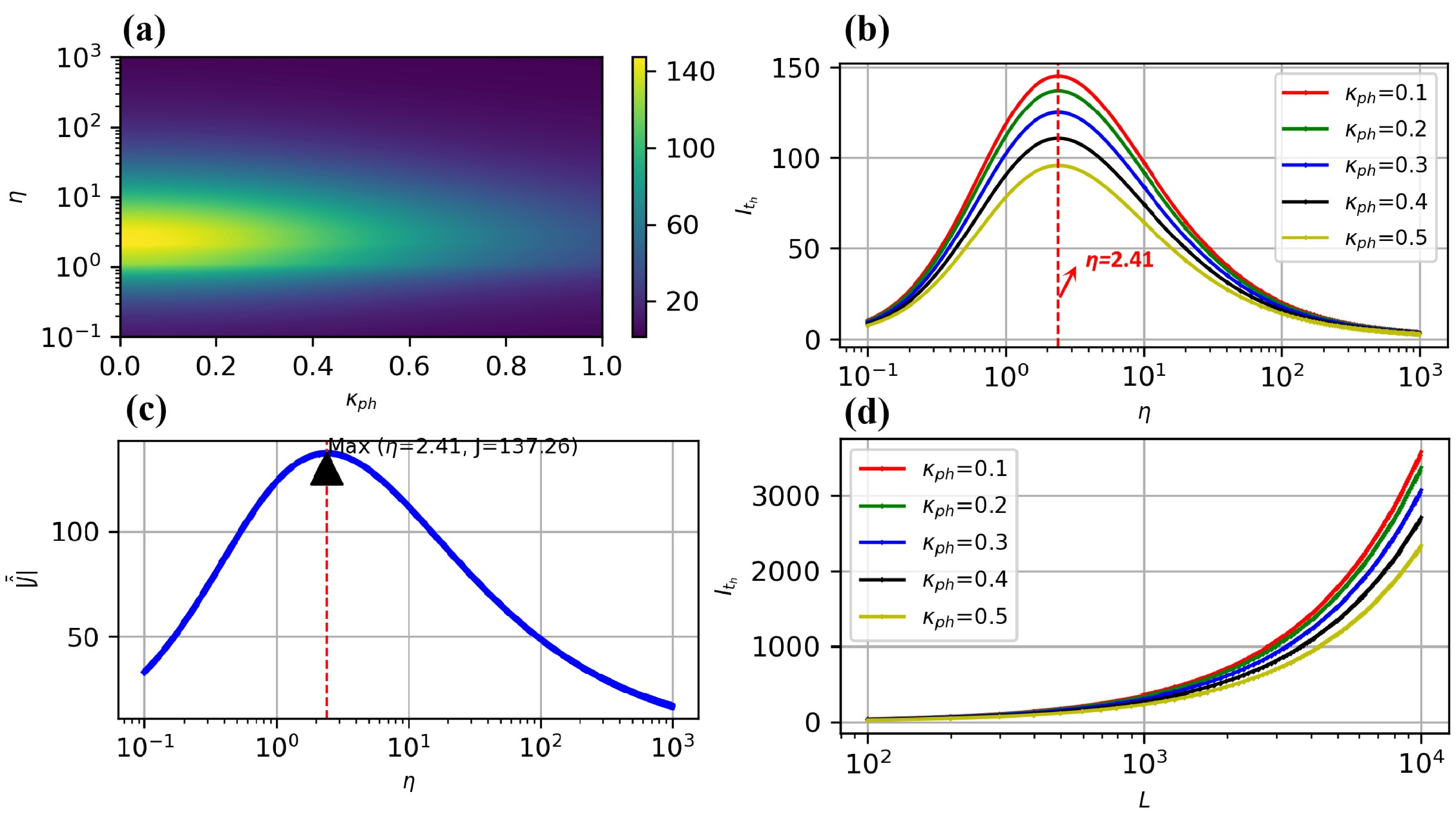}%
		\caption{(a) The QFI $\mathcal{I}_{t_h}$ as functions
			of $\eta$ and $\kappa_{{ph}}$. (b) The QFI $\mathcal{I}_{t_h}$ as functions
			of $\eta$ for some considered $\kappa_{{ph}}$ value. In (c), we show the amplitude of $\bar{\hat{J}}$ in electron steady state as a function of $\eta$. The red dashed line denotes the $\eta=2.41$. In (d), we show The QFI $\mathcal{I}_{t_h}$ as functions of the lattice site number $L$ for some considered $\kappa_{{ph}}$ value, while fix $\eta=2.41$. The coupling strength $g=0.88$, and other parameters are the same as those in Fig. 2.}
	\end{figure*}
	where $\overline{\hat{T}}(t)$ and $\overline{\hat{J}}(t)$ are the expectation values of operators $\hat{T}$ and $\hat{J}$ in the electron state. It can be also solved by a Gaussian ansatz. The dynamics are then
	fully featured by the evolution equation for the covariance matrix $\sigma$  and the non-zero  displacement vector $d$ of the state. We find the steady state by solving $\partial_{t}\sigma=0$ and $\partial_{t}d=0$. The steady state is a displaced thermal state, with $\sigma$ and $d$ are given by
	\begin{equation}
		\sigma=\begin{bmatrix}1/2&0\\0&\frac{2\Lambda^2\omega_0^2}{\left(\omega_0^2+\kappa_{ph}^2\right)^2}+1/2\end{bmatrix},d=\begin{bmatrix}-\frac{\Lambda\omega_0}{\omega_0^2+\kappa_{ph}^2}\\-\frac{\Lambda\kappa_{ph}}{\omega_0^2+\kappa_{ph}^2}\end{bmatrix},\label{16}
    \end{equation}	
    where $\Lambda = \sqrt{2L} gt_{h}(2/\eta - \frac{2(1+\eta)}{\eta \sqrt{1+2\eta}})$ with $\eta$ is the ratio of the electronic
    decay $\kappa_{el}$ and pumping rate $\Gamma$. The QFI \cite{arxiv1}  of this Gaussian state can be calculated as
    \begin{eqnarray}
    	\mathcal{I}_{t_h}&=&\frac{128}{256\mathrm{Det}^{2}\sigma-1}\bigg\{\mathrm{Det}^{2}\sigma\mathrm{Tr}[(\sigma^{-1}\dot{\sigma})^2]-\mathrm{Tr}[(\dot{\sigma}M)^2]\bigg\}\notag\\
    	&+&2\dot{d}^\mathrm{T}\sigma^{-1}\dot{d}, \label{17}
    \end{eqnarray}
    where dots denote the derivative with respect to the parameter $t_h$ and T is transposition. After some straightforward calculations, the analytical expression of the QFI for the estimation of $t_h$ is given by
    \begin{equation}
    	\mathcal{I}_{t_h}=\frac{2 \Lambda^{2}}{\mathrm{A}^{4} \mathrm{t}_{\mathrm{h}}^{2}}\left(\frac{A^{6}+4\omega_{0}^{4}\mathrm{A}^{2}\Lambda^{2}}{\mathrm{A}^{2}+4 \omega_{0}^{2} \Lambda^{2}}+\frac{256 \omega_{0}^{4} \Lambda^{2}}{16 \left(1+\frac{4 \omega_{0}^{2} \Lambda^{2}}{\mathrm{A}^{2}}\right)^{2}-1}\right), \label{18}
    \end{equation}
    One might be interested in asking what the influence of the ODC dissipative process is on the performance of quantum-metrology protocols. Is it positive or negative? To investigate this in Fig. 7(a), we illustrate the QFI $\mathcal{I}_{t_h}$ as functions of $\eta$ and $\kappa_{{ph}}$. As is evident from the figure, the cavity dissipative process will reduce the performance of quantum-metrology protocols. However, the QFI do not vary linearly with the ratio $\eta$. As is clearly described in Fig. 7(b), the maximum QFI are achieved for $\eta=2.41$ for all values of $\kappa_{{ph}}$. It is important to briefly discuss what happens in the ODC cavity system when $\eta=2.41$. As we explain in Appendix C, the expectation value of operator $\hat{T}$ is equal to 0, and the expectation value of operator $\hat{J}$ is equal to $t_{h}\sqrt{L}(2/\eta - \frac{2(1+\eta)}{\eta \sqrt{1+2\eta }})$ in the electronic steady state. Thus, the electronic kinetic energy term vanishes, leaving the non-zero current term. Meanwhile, when $\eta=2.41$, the expectation value of operator $\hat{J}$ has its maximum, as shown in Fig. 7(c). Subsequently, the quadratic term in Eq. (\ref{5}) vanishes, leaving the non-zero displaced term ($\frac{g}{\sqrt{L}}\bar{\hat{J}}\left(\hat{a}^{\dagger}+\hat{a}\right)$), which reaches its maximum when $\eta=2.41$ in the electronic steady state. Furthermore, it is worth pointing out that the size of the ODC will affect the QFI. We study this by showing the QFI $\mathcal{I}_{t_h}$ as functions of the ODC site number $L$ for some considered $\kappa_{{ph}}$ value in Fig. 7(d). As the figure shows, as $L$ increases, $\mathcal{I}_{t_h}$ experiences a sharp rise. It can be understood from Eq. \ref{19} that it consists of $L^2$, $L^1$, and $L^0$ terms. Thus, it means that an
    arbitrarily large estimation precision could, in principle, be obtained given that the size of the ODC is large enough. 
    
    Next, we calculate the CFI $F_{{t_{h}}}(\phi)$ using homodyne detection performed on the quadrature with the aim of quantifying the performance of quantum-metrology protocols in estimating hopping strength $t_h$ in the presence of ODC and cavity dissipative processes. By using Eq. \ref{7} again, one can obtain the CFI. The Wigner function of the photon state in the cavity $W[\rho_{ph}]$ has been obtained by finding the steady photon state in Appendix C. After some straightforward calculations, the analytical expression of the CFI for the estimation of $t_h$ is given by
    \begin{widetext}
    \begin{equation}
    	F_{{t_{h}}}(\phi)=\frac{2\left(16 \omega_{0}^{4} \mathrm{sin} (\phi)^{4} \Lambda^{2}+\left(\omega_{0}\mathrm{cos} (\phi)^{2}+\kappa_{{ph}} \mathrm{sin} (\phi)^{2}\right)^{2} \left(\mathrm{A}^{2}+4 \omega_{0}^{2} \mathrm{Sin} (\phi)^{2} \Lambda^{2}\right)\right) \Lambda^{2}}{\left(\mathrm{A}^{2}+4 \omega_{0}^{2} \mathrm{sin} (\phi)^{2} \Lambda^{2}\right)^{2} {t}_{{h}}^{2}}, \label{19}
    \end{equation}
    \end{widetext}
    \begin{figure*}[ht]
    	\label{Fig.8} \centering \includegraphics[width=1.0\columnwidth]{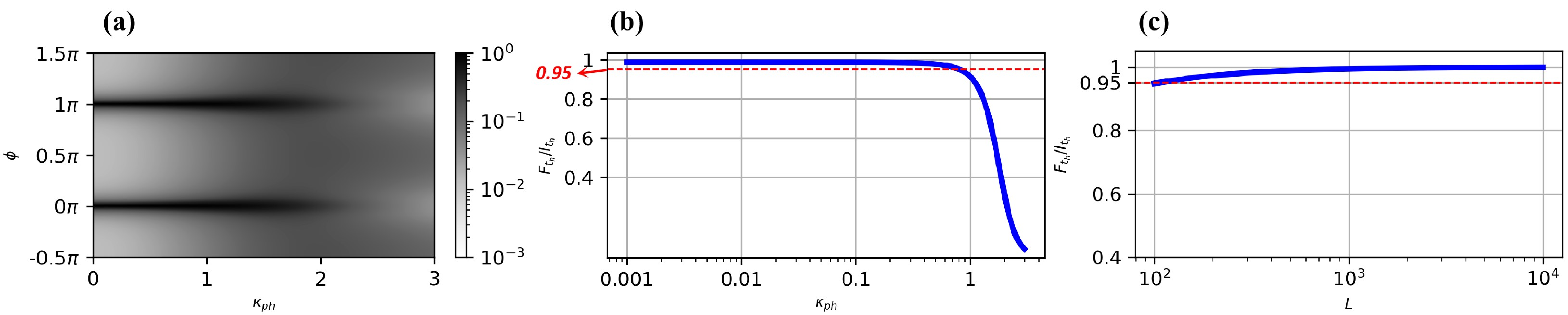}%
    	\caption{(a) $F_{{t_{h}}}(\phi)$-to-$\mathcal{I}_{t_{h}}$ ratio as functions
    		of the local oscillator phase $\phi$ and $\kappa_{{ph}}$. (b) $F_{{t_{h}}}(\phi)$-to-$\mathcal{I}_{t_{h}}$ ratio as a function
    		of $\kappa_{{ph}}$ for $\phi=0$. In (c), we show $F_{{t_{h}}}(\phi)$-to-$\mathcal{I}_{t_{h}}$ ratio as a function of the ODC site number $L$ for $\phi=0$. The red dashed line denotes the $F_{{t_{h}}}(\phi)/\mathcal{I}_{t_{h}}=0.95$. $\eta=2.41$, the coupling strength $g=0.88$, and other parameters are the same as those in Fig. 2.}
    \end{figure*}
    A crucial point in the above procedure is that the QFI can achieve their maximum value when $\eta=2.41$. Thus, it is highly desirable to find an optimal local oscillator phase $\phi$ for saturating this maximum QFI. To investigate this, in Fig. 8(a), we plot the $F_{{t_{h}}}(\phi)$-to-$\mathcal{I}_{t_{h}}$ ratio as functions of the local oscillator phase $\phi$ and $\kappa_{{ph}}$. In general, for any choice of $\kappa_{{ph}}$, the ratio is only optimal at $\phi=0$ and $\phi=\pi$.  Thus, one can still implement homodyne detection as the optimal POVM to estimate the parameter $t_h$ in the ODC cavity system in the presence of the ODC and cavity dissipative processes upon choosing $\phi=0$ or $\pi$. It is also important to see whether the ratio is optimal in the presence of a large photon decay rate. In Fig. 8(b), we plot the $F_{{t_{h}}}(\phi)$-to-$\mathcal{I}_{t_{h}}$ ratio as a function
    of $\kappa_{{ph}}$ for $\phi=0$. As the figure shows, even in the
    presence of the photon decay rate of $\kappa_{{ph}}\approx1.1$,
    $F_{{t_{h}}}(\phi)/\mathcal{I}_{t_{h}}$ shows a near-optimal ratio of 0.95. Although even for extreme photon loss of $\kappa_{{ph}}=1.2$, the ratio still reaches values up to 0.6. One might be interested to see how the $F_{{t_{h}}}(\phi)$-to-$\mathcal{I}_{t_{h}}$ ratio is affected by the ODC site number $L$. To study this in Fig. 8(c), we illustrate $F_{{t_{h}}}(\phi)/\mathcal{I}_{t_{h}}$ as a function of $L$. As the figure shows, $F_{{t_{h}}}(\phi)/\mathcal{I}_{t_{h}}$ always exhibits an optimal ratio near 1. Thus, in principle, it means that this homodyne measurement saturates the CR bound for estimating the parameter $t_h$ in the large ODC in the presence of the ODC and cavity dissipative process. 
    \section{CONCLUSION}
    \label{CONCLUSION}
    Hopping strength is an important parameter in controlling the properties of  electronic materials in condensed matter physics. Thus, in this paper, we have suggested a critical quantum metrology protocol for measuring the hopping strength of a one-dimensional chain coupled to a cavity. In this protocol, an arbitrarily large quantum estimation precision could, theoretically, be obtained near the critical point of the coupling strength in the cavity-ODC system. By implementing homodyne detection, our scheme reaches classical precision, which can completely saturate quantum precision in the ideal case quantified by QFI. Furthermore, we have shown that homodyne detection is still an optimal POVM when the cavity-ODC system operates in the dissipative regime. Remarkably, one can obtain the largest estimation precision when the electron has its maximum current in the ODC by appropriately controlling the ratio of the electron decay rate to the pumping rate in the ODC. Moreover, we find that increasing the ODC size can enhance the estimation precision in the presence of the ODC dissipative process. The results can help us better detect the changes in properties in the electronic materials depending on the hopping strength, which will pave the way for the application of quantum metrology in condensed matter cavity QED.
    \appendix
    \section{The ground state of electron}
    In the ODC-cavity interaction model, the ground state of an electron can be composed of the eigenstate of $\sum_k\hat{c}_k^{\dagger}\hat{c}_k$, with the fact that $[\sum_k\hat{c}_k^{\dagger}\hat{c}_k,H]=0$ from Eq. (\ref{2}). In the half-filled condition ($\left\langle\psi_{el}\left|\sum_k\hat{c}_k^+\hat{c}_k\right|\psi_{el}\right\rangle=L/2$), we denote the eigenstate as $\left|\psi_{el}\right\rangle=\sum_k\left|\hat{c}_k^+\hat{c}_k\right\rangle$ (FS). It is clear that $\left|\hat{c}_k^{\dagger}\hat{c}_k\right\rangle$ is the eigenstate of $\hat{c}_k^{\dagger}\hat{c}_k=\hat{n}(k)$, where $\hat{n}(k)$ is the number operator at momentum $k$. It is clear that the eigenvalue of $\hat{n}(k)=$1 or 0. According to the half-filled condition, we take as
   the ground state of electrons $\left|\psi_{el}\right\rangle$ only connected regions in k-space. Thus, the eigenvalue $n(k)$ can be written as a function of $k$
     \begin{equation}
    	n(k)=\begin{cases}0,k\not\in(s-\pi/2,s+\pi/2)\\\\1,k\in(s-\pi/2,s+\pi/2)\end{cases}, \label{A1}
    \end{equation}
    where $s$ is the center of the FS. Thus, with Eqs. (\ref3) and (\ref{A1}), the expectation values of kinetic energy and current operators in the FS can be calculated as  
    \begin{eqnarray}
    	\overline{\hat{T}}&=&\left\langle\psi_{el}\left|\hat{T}\right|\psi_{el}\right\rangle=2\frac{t_hL}\pi\mathrm{cos}(s).\notag\\
    	\overline{\hat{J}}&=&\left\langle\psi_{el}\left|\hat{J}\right|\psi_{el}\right\rangle=2\frac{t_hL}\pi\mathrm{sin}(s). \label{A2}
    \end{eqnarray}
    Thus, the effective full cavity Hamiltonian can be expressed as
    \begin{eqnarray}
    	H&=&\cos\left(\frac{g}{\sqrt{L}}(\hat{a}+\hat{a}^\dagger)\right)\overline{\hat{T}}+\sin\left(\frac{g}{\sqrt{L}}(\hat{a}+\hat{a}^\dagger)\right)\overline{\hat{J}}\notag\\
    	&&+\omega_0(\hat{a}^\dagger \hat{a}+\frac{1}{2}).\label{A3}
    \end{eqnarray} 
    We have shown the spectrum of the full cavity Hamiltonian in Fig. 2(a) and found that the FS is always located at the FS centered around $s = 0$ for all considered coupling strengths. Thus, in this electronic ground state, the expectation value of the current density $\bar{\hat{J}}=0$ and the kinetic energy operator $\bar{\hat{T}}=2\frac{t_hL}\pi$. Notice that these results are still valid when the cavity Hamiltonian is truncated at the second-order as $g<g_{c}$.
    \section{Cavity dissipative dynamics}
    We will now describe the dynamics of the system in the presence of a cavity dissipative process under the Lindblad equation: $\partial_t\hat{\rho}=-i{\left[H^{2^{\mathrm{nd}}},\hat{\rho}\right]}+\kappa_{ph}\mathcal{L}{\left[\hat{a}\right]}\hat{\rho}$, given in the main text. The density matrix of the cavity can be obtained by $\hat{\rho}_{ph}=\mathrm{Tr_{el}}\{\hat{\rho}\}$ where $\mathrm{Tr_{el}}\{\}$ is defined as the trace over the electron degrees of freedom with $\hat{\rho}=\hat\rho_{ph}\otimes\hat\rho_{el}$. Thus, we have 
    \begin{equation}
    	\partial_t\hat{\rho}_{ph}=-i{\left\lfloor\mathrm{Tr_{el}}\{H^{2^{\mathrm{nd}}}\hat{\rho}_{el}\},\hat{\rho}_{ph}\right\rfloor}+\mathrm{Tr_{el}}\{\hat{\rho}_{el}\}\otimes\kappa_{ph}\mathcal{L}{\left[\hat{a}\right]}\hat{\rho}_{ph},\label{B1}
    \end{equation} 
    where $\mathrm{Tr_{el}}\{H^{2^{\mathrm{nd}}}\hat{\rho}_{el}\}$ is the expectation value of $H^{2^{\mathrm{nd}}}$ in the electronic ground state. Thus, operator $\hat{T}$ and $\hat{J}$ can be replaced by $\overline{\hat{T}}$ and $\overline{\hat{J}}$ in Eq. (\ref{A2}) with $s=0$.  Notice that they remain unchanged in the ground state of electron due to the fact that $[\hat{c}_k^\dagger \hat{c}_k,H^{2^{\mathrm{nd}}}]=0$. Thus, we have
    \begin{equation}
    	\partial_{t}\hat{\rho}_{ph}=-i\Bigg[ \omega_{0}\hat{a}^{\dagger}\hat{a}-\omega_{0}X \frac{g^{2}}{4}(\hat{a}+\hat{a}^{\dagger})^{2},\hat{\rho}_{ph}\Bigg]+\kappa_{ph}\mathcal{L}[\hat{a}]\hat{\rho}_{ph}, \label{B2}
    \end{equation} 
    where $X=\frac{4t_h}{\omega_0\pi}$. We will now move to phase space \cite{arxiv2} and rewrite the Lindblad equation above into a Fokker-Planck equation for the Wigner function:
    \begin{widetext}
    	\begin{equation}
    \frac{\partial W\left(x,p\right)}{\partial t}=-\omega_{0}p\frac{\partial W}{\partial x}+\omega_{0}(1-X\mathrm{g}^{2})x\frac{\partial W}{\partial p}+\kappa_{{_{ph}}}(2W+\sum_{i=1}^{2}x_{i}\partial_{i}W+\sum_{i,j=1}^{2}\partial_{i}\sigma_{ij}^{Lin}\partial_{j}W),\label{B3}
    \end{equation}
    	\end{widetext}
    where $x_{1}=x,x_{2}=p,$ and $\sigma_{ij}^{ Lin}=\begin{bmatrix}1&0\\0&1\end{bmatrix}/2$. This equation can be solved by a Gaussian ansatz $W=\frac{1}{2\pi\sqrt{\mathrm{det}(\sigma)}}\exp\big\{\frac{-1}{2}\left(x_{i}-d_{i})(\sigma^{-1}\right)_{ij}\left(x_j-d_j\right)\big\}$ where $\sigma_{11}=\left\langle x^2\right\rangle-\left\langle x\right\rangle^2,\sigma_{22}=\left\langle p^2\right\rangle-\left\langle p\right\rangle^2$ and $\sigma_{12}=\sigma_{21}=\left\langle xp\right\rangle/2-\left\langle x\right\rangle\left\langle p\right\rangle$. $d_i$ and $d_j$ are the elements of the displacement $\overline{d}=\begin{bmatrix}\overline{x_i}\\\overline{x_j}\end{bmatrix}$ which decay at a rate $2\kappa_{{ph}}$ and will quickly reach 0. Therefore, the
    covariance matrix, which fully characterizes this function, is outlined in the following equation
    \begin{equation}
    \partial_t\sigma=B\sigma+\sigma B^T-2\kappa_{ph}(\sigma-\sigma^{Lin}),\label{B4}
    \end{equation}
    where $B=\begin{bmatrix}0&\omega_0\\\omega_0(Xg^2-1)&0\end{bmatrix}$. Finally, we find the steady state by solving $\partial_{t}\sigma=0$, and $\sigma$ is given by
    	\begin{equation}
    		\sigma=\begin{bmatrix}\frac{2\kappa_{ph}^2+\omega_0^2-\omega_0^2(Xg^2-1)}{4(\kappa_{ph}^2-\omega_0^2(Xg^2-1))}&\frac{Xg^2\kappa_{ph}\omega_0}{4(\kappa_{ph}^2-\omega_0^2(Xg^2-1))}\\\frac{Xg^2\kappa_{ph}\omega_0}{4(\kappa_{ph}^2-\omega_0^2(Xg^2-1))}&\frac{2\kappa_{ph}^2+\omega_0^2(Xg^2-1)^2-\omega_0^2(Xg^2-1)}{4(\kappa_{ph}^2-\omega_0^2(Xg^2-1))}\end{bmatrix}. \label{B5}
    	\end{equation} 
    \section{The steady state of electron}
    Here we briefly study the steady state of electron in the presence of ODC dissipative process, which can be described by jump operators $\hat{L}_j=\sqrt{\kappa_{el}}(\hat{c}_j-i\hat{c}_{j+1})$ and $\hat{G}_{j}=\sqrt{2\Gamma}\hat{c}_{j}^{\dagger}$, where $\kappa_{el}$ is the
    electron decay rate and $\Gamma$ is the electron pumping rate. Notice that we assume that the cavity is not coupled to the ODC dissipative process. We can write the Lindblad equation in momentum space:
    \begin{eqnarray}
    		\partial_{t}\hat{\rho}&=&-i{\left\lfloor H^{2^{\mathrm{nd}}},\hat{\rho}\right\rfloor}+\kappa_{ph}L{\left[\hat{a}\right]}\hat{\rho}\notag\\
    		&-&2\kappa_{el}\left(1+\sin(k)\right)\sum_{k}\mathcal{D}{\left[\hat{c}_k\right]}\hat{\rho}\notag\\
    		&-&2\Gamma\sum_{k}\mathcal{D}{\left[\hat{c}_k^{\dagger}\right]}\hat{\rho},\label{C1}
    \end{eqnarray}
    where $\mathcal{D}[\hat{O}]\hat{X}=2\hat{O}\hat{X}\hat{O}^{\dagger}-\hat{X}\hat{O}^{\dagger}\hat{O}-\hat{O}^{\dagger}\hat{O}\hat{X}$. Next, we will calculate  the expectation value of the number operator $\hat{c}_k^{\dagger}\hat{c}_k=\hat{n}(k)$, which is entirely characterized  by the equation
    \begin{widetext}
    	\begin{eqnarray}
    		\mathrm{Tr}\left\{\hat{n}_k\partial_t\left(\hat{\rho}_{ph}\otimes\hat{\rho}_{el}\right)\right\}&=&-i\mathrm{Tr}\{\hat{n}_k\left[H^{2^\mathrm{nd}},\hat{\rho}_{ph}\otimes\hat{\rho}_{el}\right]+\kappa_{ph}\mathrm{Tr}(L[\hat{a}]\hat{\rho}_{ph}\otimes\hat{n}_k\hat{\rho}_{el})\notag\\
    		&+&2\kappa_{el}\left(1+\sin(k)\right)\sum_k\mathrm{Tr}(\hat{n}_kD[\hat{c}_k]\hat{\rho}_{el}\otimes\hat{\rho}_{ph})+2\Gamma\sum_k\mathrm{Tr}(\hat{n}_kD[\hat{c}_k^{\dagger}]\hat{\rho}_{el}\otimes\hat{\rho}_{ph}),\label{C2}
    	\end{eqnarray}
    \end{widetext}
    where $\mathrm{Tr}\{\}$ is defined as the trace over the electron and photon degrees of freedom, while we find term $\mathrm{Tr}\{\hat{n}_{k}\left[H^{2^{\mathrm{nd}}},\hat{\rho}_{ph}\otimes\hat{\rho}_{el}\right]=0$ and $\kappa_{ph}\mathrm{Tr}(L[\hat{a}]\hat{\rho}_{ph}\otimes\hat{n}_k\hat{\rho}_{el})=0$. Thus, the equation can be reduced to
    \begin{widetext}
    	\begin{equation}
    		\mathrm{Tr}\left\{\hat{n}_{k}\partial_{t}\left(\hat{\rho}_{ph}\otimes\hat{\rho}_{el}\right)\right\}=2\kappa_{el}\left(1+\sin(k)\right)\sum_{k}\mathrm{Tr}(\hat{n}_{k}D[\hat{c}_{k}]\hat{\rho}_{el}\otimes\hat{\rho}_{ph})+2\Gamma\sum_{k}\mathrm{Tr}(\hat{n}_{k}D[\hat{c}_{k}^{\dagger}]\hat{\rho}_{el}\otimes\hat{\rho}_{ph}).\label{C3}
    	\end{equation}
    \end{widetext}
    We find the steady state by solving $\mathrm{Tr}\left\{\hat{n}_{k}\partial_{t}\left(\hat{\rho}_{ph}\otimes\hat{\rho}_{el}\right)\right\}=0$, the solution is 
    \begin{equation}
    	\langle\hat{n}_{k}\rangle_{\mathrm{ss}}=\frac{1}{e^{\beta\epsilon(k)}+1},\label{C4}
    \end{equation}
    where $e^{\beta\varepsilon(k)}=\eta(1+\sin(k))$ with $\eta=\kappa_{el}/\Gamma $. Thus, in the electron steady state, we have
    \begin{eqnarray}
    \overline{\hat{T}}(\infty)&=&\sum_k-2t_h\cos(k)\langle\hat{n}_{k}\rangle_{\mathrm{ss}}=0,\notag\\
    \overline{\hat{J}}(\infty)&=&\sum_k2t_h\sin(k)\langle\hat{n}_{k}\rangle_{\mathrm{ss}}={\Lambda\sqrt{L}} /({g\sqrt{2}}),\label{C5}
    \end{eqnarray} 
    where $\overline{\hat{T}}(\infty)$ and $\overline{\hat{J}}(\infty)$ are the expectation values of kinetic energy and current operators at $t=\infty$.
    \section{ODC dissipative dynamics}
    Here we briefly study the dynamics of the photon field in the presence of the ODC dissipative process. Next, we eliminate the electron mode by taking the trace over the electron degrees of freedom in Eq. (\ref{C1}). Thus, we have 
    \begin{widetext}
    	\begin{eqnarray}
    		\partial_{{_{t}}}\hat{\rho}_{{_{ph}}}&=&-i{\left\lfloor\mathrm{Tr_{el}}\{H^{{2^{{nd}}}}\hat{\rho}_{{_{el}}}\},\hat{\rho}_{{_{ph}}}\right\rfloor}+\mathrm{Tr}_{\mathrm{el}}\{\kappa_{ph}\mathcal{L}[\hat{a}]\hat{\rho}_{ph}\otimes\hat{\rho}_{el}\}\notag\\
    		&-&2\kappa_{{_{el}}}(1+\sin(k))\sum_{k}\mathrm{Tr_{el}}\{{\mathcal{D}[\hat{c}_{{_{k}}}]\hat{\rho}_{{_{el}}}}\}\otimes\hat{\rho}_{{_{ph}}}
    		-2\Gamma\sum_{k}\mathrm{Tr_{el}}\{\mathcal{D}[\hat{c}_{{_{k}}}^{\dagger}]\hat{\rho}_{{_{el}}}\}\otimes\hat{\rho}_{{_{ph}}},\label{D1}
    	\end{eqnarray}
    \end{widetext}
    where we find terms $\mathrm{Tr}\{{\mathcal{D}[\hat{c}_{{_{k}}}]\hat{\rho}_{{_{el}}}}\}$ and $\mathrm{Tr}\{\mathcal{D}[\hat{c}_{{_{k}}}^{\dagger}]\hat{\rho}_{{_{el}}}\}$ are equal to 0. Thus, the equation can be reduced to
    \begin{equation}
    	\partial_t\hat{\rho}_{ph}=-i\lfloor H^{2^{\mathrm{nd}}}(\overline{\hat{T}}(t),\overline{\hat{J}}(t)),\hat{\rho}_{ph}\rfloor+\kappa_{ph}\mathcal{L}[\hat{a}]\hat{\rho}_{ph}.\label{D2}
    \end{equation}
    We can rewrite the Lindblad equation above into a Fokker-Planck equation for the Wigner function:
    \begin{widetext}
    \begin{equation}
    \frac{\partial W\left(x,p\right)}{\partial t}=-\omega_0p\frac{\partial W}{\partial x}+\{\omega_0-\frac{2g^2}{L}\overline{\hat{T}}\left(t\right)\}x\frac{\partial W}{\partial p}+\frac{g}{\sqrt{L}}\overline{\hat{J}}\left(t\right)\frac{\partial W}{\partial p}+\kappa_{ph}\left(2W+\sum_{i=1}^2x_i\partial_iW+\sum_{i,j=1}^2\partial_i\sigma_{ij}^{Lin}\partial_jW\right).\label{D3}		
    \end{equation}
    \end{widetext}
    This equation can also be solved by a Gaussian ansatz $W=\frac{1}{2\pi\sqrt{\mathrm{det}(\sigma)}}\exp\big\{\frac{-1}{2}\left(x_{i}-d_{i})(\sigma^{-1}\right)_{ij}\left(x_j-d_j\right)\big\}$. Thus, the photon steady state can be obtained by solving $ {\partial W\left(x,p\right)}/{\partial t}=0$ with $\overline{\hat{T}}(\infty)$ and $\overline{\hat{J}}(\infty)$ in Eq .(\ref{C5}). Thus, Eq. (\ref{D3}) can be simplified as
    \begin{widetext}
    \begin{equation} 
    -\omega_0p \frac{\partial W}{\partial x}+\omega_0x \frac{\partial W}{\partial p}+\Lambda \frac{\partial W}{\partial p}+\kappa_{ph}(2W+\sum_{i=1}^2x_i\partial_iW+\sum_{i,j=1}^2\partial_i\sigma_{ij}^{Lin}\partial_jW)=0. \label{D4}				
    \end{equation}
    \end{widetext}
    According to this equation, we find	the following equation, which characterizes the photon steady state
    \begin{equation}
    \begin{cases}\omega_0\left\langle p\right\rangle-\kappa_{ph}\left\langle    x\right\rangle=0,\\\omega_0\left\langle x\right\rangle+\Lambda+\kappa_{ph}\left\langle p\right\rangle=0,\\2\omega_0\left\langle xp\right\rangle-2\kappa_{ph}\left\langle x^2\right\rangle+\kappa_{ph}=0,\\2\omega_0\left\langle xp\right\rangle+2\Lambda\left\langle p\right\rangle+2\kappa_{ph}\left\langle p^2\right\rangle-\kappa_{ph}=0,\\\omega_0\left\langle p^2\right\rangle-\omega_0\left\langle x^2\right\rangle-\Lambda\left\langle x\right\rangle-2\kappa_{ph}\left\langle xp\right\rangle=0,\end{cases}	\label{D5}
    \end{equation}
    Solving this equation, we find $\langle xp\rangle=\Lambda^{2}\omega_{0}\kappa_{pb}/(\omega_{0}^{2}+\kappa_{pb}^{2})^{2},\left\langle x^{2}\right\rangle=\Lambda^{2}\omega_{0}^{2}/(\omega_{0}^{2}+\kappa_{pb}^{2})^{2}+1/2,\left\langle p^{2}\right\rangle=\Lambda^{2}\omega_{0}^{2}/(\omega_{0}^{2}+\kappa_{pb}^{2})^{2}+\Lambda^{2}/(\omega_{0}^{2}+\kappa_{pb}^{2})+1/2,\left\langle p\right\rangle=\Lambda\kappa_{pb}/(\omega_{0}^{2}+\kappa_{pb}^{2})$ and $\left\langle x\right\rangle=-\Lambda\omega_{0}/(\omega_{0}^{2}+\kappa_{pb}^{2})$.
    Finally, we obtain the displacement and covariance matrix
    \begin{equation}
    	\overline{d}=\begin{bmatrix}-\frac{\Lambda\omega_0}{\omega_0^2+\kappa_{ph}^2}\\-\frac{\Lambda\kappa_{ph}}{\omega_0^2+\kappa_{ph}^2}\end{bmatrix},\sigma=\begin{bmatrix}1/2&0\\0&\frac{2\Lambda^2\omega_0^2}{\left(\omega_0^2+\kappa_{ph}^2\right)^2}+1/2\end{bmatrix}. \label{D7}
    \end{equation}
	\bibliography{apssamp}
	\end{document}